\documentclass[%
superscriptaddress,
preprint,
 amsmath,amssymb,
 aps,
prstab,
]{revtex4-2}

\usepackage{graphicx}
\usepackage{subfigure}
\usepackage{dcolumn}
\usepackage{bm}
\usepackage{multirow,ragged2e}

\usepackage{xcolor} 
\usepackage{mathtools} 

\date{\today}

\begin{document}

\title{General formulation of impedance in Frenet-Serret coordinate system}

\author{Demin Zhou}
 \altaffiliation[Also at ]{Graduate University for Advanced Studies (SOKENDAI)}
 \email{dmzhou@post.kek.jp}
\affiliation{%
KEK, High Energy Accelerator Organization, Oho 1-1, Tsukuba 305-0801, Japan
}%
\author{Cheng-Ying Tsai}%
 \affiliation{School of Electrical and Electronic Engineering, Huazhong University of Science and Technology, Wuhan 430074, China}

\date{\today}

\begin{abstract}
In accelerator physics, the concept of impedance is popularly used to describe the interactions of charged particles inside a bunch or between bunches in a train. Standard formulations of impedance assume that the driving charge has a constant velocity $\vec{v}=v\vec{i}_z$ in the $z$ direction of the Cartesian coordinate system. For the case of driving charge moving along a curved orbit, impedance can be formulated in the Frenet-Serret coordinate system, but there seems to be a lack of systematic formulations. This note presents an effort in this direction.
\end{abstract}

\maketitle

\section{Introduction}

The problem is related to finding the solution of Maxwell's equations with the trajectory of the specified charged driving beam. In accelerator physics, in principle, studying the dynamics of charged beams requires solving the coupled Lorentz-Maxwell equations. The charged particles are assumed to have zero sizes. Mathematically, the Dirac delta function is used to describe the positions of the particles. The charge and current density are derived from the spatial distribution.

To solve an electromagnetic problem, one may start by finding the solution in one inertial frame and then apply the relativistic transformation to find the solution in another inertial frame. In the case of charged particles' motion under guiding fields, the particles are accelerated. Therefore, it is not trivial to apply this approach.


\section{Source charge: Trajectory, charge and current densities}

The coordinates of a particle in the Frenet-Serret (F-S) coordinate system can be written as
\begin{equation}
    \vec{r}(s)=x(s)\Vec{e}_x(s) + y(s)\Vec{e}_y(s) + \Vec{r}_0(s),
    \label{eq:ParticleCoordinates_FS}
\end{equation}
with $\vec{r}_0(s)$ the trajectory of an ideal particle, and $s$ the path length along the trajectory. In fact, $s$ and $\vec{r}_0(s)=\vec{r}_0(s(t))$ are functions of time. Given the velocity vector $\Dot{\vec{r}}_0(t)=d\vec{r}_0(t)/dt$, $s$ is defined as
\begin{equation}
    s(t)=\int_0^t \lVert \Dot{\vec{r}}_0(t') \rVert  dt'.
    \label{eq:Path_Length_Definition}
\end{equation}
Though the trajectory of the ideal particle $\vec{r}_0(s)$ is prescribed with $s$ given by Eq.(\ref{eq:Path_Length_Definition}), any particle with its coordinates $\vec{r}(s)$ given by Eq.(\ref{eq:ParticleCoordinates_FS}) has its own $s$. This can be interpreted as follows. Given the particle's coordinates $\vec{r}(t)=(X(t),Y(t),Z(t))$ and a prescribed curve $\vec{r}_0(t)$ in the Cartesian coordinate system, in the F-S coordinate system with reference to $\vec{r}_0(t)$, the particle's coordinates are $\vec{r}(t)=(x(t),y(t),s(t))$. Equation (\ref{eq:Path_Length_Definition}) enables us to choose $s$ as an independent variable. Then the time $t$ will be a variable to determine the position of the particle.

We assume no acceleration; the particle under consideration has a constant velocity $v=dr/dt$. With the path length as the independent variable, to describe the particle's motion, the derivative over $t$ (in Cartesian coordinate system and lab frame) can be replaced by (see~\cite{wiedemann2015particle}, page 110)
\begin{equation}
    \frac{d}{dt}=v\frac{d}{dr}=v\frac{ds}{dr} \frac{d}{ds}=\frac{v}{r'}\frac{d}{ds},
    \label{eq:Time_Derivative_to_Path_Derivative}
\end{equation}
with $r'=dr/ds$ to be explicitly formulated and the prime indicating the derivative over $s$ in the whole of this paper.



The vector $\vec{r}_0(s(t))$ defines the F-S coordinate system along the reference trajectory of an ideal particle. This is how to define the F-S coordinates.
\begin{equation}
    \frac{d\vec{r}_0(s(t))}{dt} \equiv
    \frac{d\vec{r}_0(s)}{ds} \frac{ds}{dt}
    =v_0 \vec{e}_s.
\end{equation}
This is justified by $r'=dr/ds=1$ in Eq.(\ref{eq:Time_Derivative_to_Path_Derivative}). We define the unit vector tangent to the ideal particle's trajectory as
\begin{equation}
    \vec{e}_s \equiv \frac{d\vec{r}_0(s)}{ds},
\end{equation}
and constant scalar velocity of the ideal particle as
\begin{equation}
    v_0 \equiv \left . \frac{ds}{dt} \right |_{x=y=0}.
\end{equation}
$v_0$ is the velocity of the ideal particle moving along the reference trajectory and is assumed to be constant. For simplification, we only consider the curved reference trajectory in the horizontal plane. The relations of the right-handed orthonormal basis are
\begin{equation}
    \frac{d\vec{e}_x}{ds}=\frac{1}{\rho_x(s)}\vec{e}_s,
\end{equation}
\begin{equation}
    \frac{d\vec{e}_y}{ds}=0,
\end{equation}
\begin{equation}
    \frac{d\vec{e}_s}{ds}=-\frac{1}{\rho_x(s)}\vec{e}_x.
\end{equation}
The curvature of the reference trajectory is determined by $\rho_x(s)$ which is assumed to be $s$-dependent.

In general, using $s$ as the independent variable, the velocity of the source particle is
\begin{equation}
    \vec{v}= 
    \frac{d\vec{r}(t)}{dt}
    = \frac{ds}{dt} (g\vec{e}_s+x'\vec{e}_x+y'\vec{e}_y),
    \label{eq:Velocity_in_FS_frame_General}
\end{equation}
with $g=1+x/\rho_x(s)$, $x'=dx/ds$ and $y'=dy/ds$. Note that $ds/dt$ should be a function of $(x,y,s)$ (following Eq.(A.13) of~\cite{yokoya1985beam}):
\begin{equation}
    \frac{ds}{dt}=
    \frac{v}{\sqrt{g^2+x'^2+y'^2}},
    \label{eq:Arclength_Time_relation}
\end{equation}
with $v=|\vec{v}|$. When using $s$ as the independent variable, this equation also determines the arrival time with given $s$ along the path of the reference trajectory.

Here we mainly consider a special case in which $x_1$ and $y_1$ are constants (i.e., they are independent of time and spatial positions: $x_1=y_1=constant$ and $x_1' = y_1' = 0$)
\begin{equation}
    \vec{r}_1(s)=x_1\Vec{e}_x + y_1\Vec{e}_y + \Vec{r}_0(s).
\end{equation}
For this case
, there is (following Eq. (26) of~\cite{li2008curvature})
\begin{equation}
    \left . \frac{ds}{dt} \right |_{x_1,y_1} 
    = \frac{1}{g_1} v_0,
\end{equation}
with $g_1=1+x_1/\rho_x(s)$. This ensures that the velocity of an arbitrary source particle in the lab frame is the same as that of the reference particle,
\begin{equation}
    \vec{v}_1 = 
    \frac{d\vec{r}_1(t)}{dt}
    =v_0\vec{e}_s.
    \label{eq:Velocity_in_FS_frame}
\end{equation}

In F-S coordinates, the source-charge density is defined as
\begin{equation}
    \rho_1(\vec{r},s) = q_1\delta(\vec{r}-\vec{r}_1(t(s)))
    = \frac{q_1}{g} \delta(x-x_1) \delta(y-y_1) \delta(s-s_1),
    \label{eq:Source_charge_density_of_point_charge_time_domain}
\end{equation}
with $s_1=v_0t$. Note that here $s$ represents the path length measured along the trajectory of the ideal particle and $t$ represents the time taken by the ideal particle along its prescribed path. The factor $g=1+x/\rho(s)$ appears due to the definition of the Dirac delta function in the F-S coordinate system. It ensures that the conservation law will be obeyed in the F-S coordinate system. Also, note that the volume element in the Cartesian system is $dV=dxdydz$ and in the F-S system is $dV=gdxdyds$.
Consequently, the source-current density (It indeed satisfies the continuity equation) is
\begin{equation}
    \vec{j}_1(\vec{r},s)
    = q_1v_0\vec{e}_s \delta(x-x_1) \delta(y-y_1) \delta(s-v_0t).
    \label{eq:Source_current_density_of_point_charge_time_domain}
\end{equation}
Note that the relation derived in Cartesian system $\vec{j}_1(\vec{r},t)=\rho_1(\vec{r},t) \vec{v}_1$ does not apply here.

With the above definitions of charge and current densities, we can see that a source particle with $x\neq 0$ will have the same absolute velocity as that of the reference particle. This guarantees that our model is physical. Therefore, the rigid beam approximation in the F-S frame for distribution in the $x$-direction is physical.

In~\cite{bizzozero2015studies}, Bizzozero claimed that the rigid charge density in the form of $\rho(\vec{r},t)=\lambda(s-\beta ct)w(x)h(y)$ is not physical if $w(x)$ is a continuous distribution. The spread in $x$ creates a velocity dispersion over the source: Particles' velocity inside the beam varies with $x$. It is suggested to choose $w(x)=\delta(x)$ so that the charge density can maintain a rigid profile without velocity dispersion.
However, according to our formulations, there is no problem of velocity dispersion at all. The essential point is that in the Frenet-Serret space, we have to define the charge and current densities as Eqs.(\ref{eq:Source_charge_density_of_point_charge_time_domain}) and (\ref{eq:Source_current_density_of_point_charge_time_domain}).

We define the Fourier transform pair as
\begin{equation}
    f(\vec{r},\omega)=\int_{-\infty}^\infty f(\vec{r},t)e^{i\omega t}dt,
\end{equation}
\begin{equation}
    f(\vec{r},t)=\frac{1}{2\pi}\int_{-\infty}^\infty f(\vec{r},\omega)e^{-i\omega t}d\omega.
\end{equation}

Then the spectra of charge and current densities are
\begin{equation}
    \rho_1(\vec{r},\omega)=
    \frac{q_1}{g_1v_0}\delta(x-x_1) \delta(y-y_1) e^{i\omega \frac{s}{v_0}},
    \label{eq:Source_charge_density_freq_domain}
\end{equation}
\begin{equation}
    \vec{j}_1(\vec{r},\omega)=
    q_1\vec{e}_s\delta(x-x_1) \delta(y-y_1) e^{i\omega \frac{s}{v_0}}.
    \label{eq:Source_current_density_freq_domain}
\end{equation}

\section{Maxwell's equations}

Maxwell's equations are
\begin{equation}
    \vec{\nabla}\times \vec{E}(\vec{r},t) = 
    -\frac{\partial \vec{B}(\vec{r},t)}{\partial t},
\end{equation}
\begin{equation}
    \vec{\nabla} \times \vec{B}(\vec{r},t) 
    -\frac{1}{c^2} \frac{\partial \vec{E}(\vec{r},t)}{\partial t}
    = \mu_0 \vec{j}(\vec{r},t),
\end{equation}
\begin{equation}
    \vec{\nabla}\cdot \vec{B}(\vec{r},t) = 0,
\end{equation}
\begin{equation}
    \vec{\nabla}\cdot \vec{E}(\vec{r},t) = 
    \frac{\rho(\vec{r},t)}{\epsilon_0}.
\end{equation}
The continuity equation is derived from Maxwell's equations
\begin{equation}
    \vec{\nabla}\cdot \vec{j}(\vec{r},t) = 
    -\frac{\partial \rho(\vec{r},t)}{\partial t}.
\end{equation}
This equation implies local conservation of charge. A local conservation law for energy in the electromagnetic field (Poynting’s theorem on conservatin of energy) can be derived from Maxwell's equations:
\begin{equation}
    \vec{\nabla} \cdot \vec{S} 
    + \frac{\partial u}{\partial t} 
    + \vec{E} \cdot \vec{j} = 0,
    \label{eq:Poynting_theorem}
\end{equation}
where $\vec{S}=\frac{1}{\mu_0} \vec{E}\times \vec{B}$ is the Poynting's vector (it represents the energy flux in the field), $u=\frac{\epsilon_0}{2}E^2+\frac{1}{2\mu_0}B^2$ is the energy density in the field, and $\vec{E}\cdot \vec{j}$ represents the rate per unit volume of energy loss from the EM field to the space (in our case, it is vacuum), or energy gain from the environment to the beam if the sign is negative. Equation (\ref{eq:Poynting_theorem}) is derived from
\begin{equation}
    \begin{split}
        & \vec{B}\cdot \left( \vec{\nabla} \times \vec{E} 
        + \frac{\partial \vec{B}}{\partial t}\right)
        - \vec{E}\cdot \left( \vec{\nabla} \times \vec{B} 
        - \frac{1}{c^2} \frac{\partial \vec{E}}{\partial t} -\mu_0\vec{j}\right)
        \\
        &= \left[
        \vec{B}\cdot \vec{\nabla} \times \vec{E}
        - \vec{E}\cdot \vec{\nabla} \times \vec{B}
        \right]
        +
        \left[
        \vec{B}\cdot \frac{\partial \vec{B}}{\partial t}
        + \frac{1}{c^2} \vec{E}\cdot \frac{\partial \vec{E}}{\partial t}
        \right]
        +
        \mu_0\vec{j} \cdot \vec{E}
        =0.
    \end{split}
\end{equation}
Note that $\vec{\nabla}\cdot (\vec{E}\times\vec{B})=\vec{B}\cdot \vec{\nabla} \times \vec{E} - \vec{E}\cdot \vec{\nabla} \times \vec{B}$.

Going to the frequency domain, we have
\begin{equation}
    \vec{\nabla}\times \vec{E}(\vec{r},\omega) = 
    i\omega \vec{B}(\vec{r},\omega),
\end{equation}
\begin{equation}
    \vec{\nabla} \times \vec{B}(\vec{r},t) 
    +\frac{i\omega}{c^2} \vec{E}(\vec{r},\omega)
    = \mu_0 \vec{j}(\vec{r},\omega),
\end{equation}
\begin{equation}
    \vec{\nabla}\cdot \vec{B}(\vec{r},\omega) = 0,
\end{equation}
\begin{equation}
    \vec{\nabla}\cdot \vec{E}(\vec{r},\omega) = 
    \frac{\rho(\vec{r},\omega)}{\epsilon_0},
\end{equation}
\begin{equation}
    \vec{\nabla}\cdot \vec{j}(\vec{r},\omega) = 
    i\omega\rho(\vec{r},\omega).
    \label{eq:Continuity_equation_freq_domain}
\end{equation}
It can be checked that Eqs.(\ref{eq:Source_charge_density_freq_domain}) and (\ref{eq:Source_current_density_freq_domain}) satisfy Eq.(\ref{eq:Continuity_equation_freq_domain}) in the F-S coordinate system, just as Eqs.(13,14) satisfy Eq.(23).

The explicit forms of the frequency-domain Maxwell's equations in F-S coordinates are
\begin{equation}
    \frac{1}{g} \left[
    \frac{\partial}{\partial y} \left( gE_s(\vec{r},\omega) \right)
    - \frac{\partial}{\partial s} E_y(\vec{r},\omega)
    \right]
    = i\omega B_x(\vec{r},\omega),
\end{equation}
\begin{equation}
    \frac{1}{g} \left[
    \frac{\partial}{\partial s} E_x(\vec{r},\omega)
    - \frac{\partial}{\partial x} \left( gE_s(\vec{r},\omega) \right)
    \right]
    = i\omega B_y(\vec{r},\omega),
\end{equation}
\begin{equation}
    \frac{\partial}{\partial x} E_y(\vec{r},\omega)
    - \frac{\partial}{\partial y} E_x(\vec{r},\omega)
    =i\omega B_s(\vec{r},\omega),
\end{equation}

\begin{equation}
    \frac{1}{g} \left[
    \frac{\partial}{\partial y} \left( gB_s(\vec{r},\omega) \right)
    - \frac{\partial}{\partial s} B_y(\vec{r},\omega)
    \right]
    + \frac{i\omega}{c^2} E_x(\vec{r},\omega)
    = \mu_0 j_x(\vec{r},\omega),
\end{equation}
\begin{equation}
    \frac{1}{g} \left[
    \frac{\partial}{\partial s} B_x(\vec{r},\omega)
    - \frac{\partial}{\partial x} \left( gB_s(\vec{r},\omega) \right)
    \right]
    + \frac{i\omega}{c^2} E_y(\vec{r},\omega)
    = \mu_0 j_y(\vec{r},\omega),
\end{equation}
\begin{equation}
    \frac{\partial}{\partial x} B_y(\vec{r},\omega)
    - \frac{\partial}{\partial y} B_x(\vec{r},\omega)
    + \frac{i\omega}{c^2} E_s(\vec{r},\omega)
    =\mu_0 j_s(\vec{r},\omega),
\end{equation}

\begin{equation}
    \frac{1}{g} \left[
    \frac{\partial}{\partial x} \left( gB_x(\vec{r},\omega) \right)
    + \frac{\partial}{\partial y} \left( gB_y(\vec{r},\omega) \right)
    +\frac{\partial}{\partial s} B_s(\vec{r},\omega)
    \right]
    = 0,
\end{equation}

\begin{equation}
    \frac{1}{g} \left[
    \frac{\partial}{\partial x} \left( gE_x(\vec{r},\omega) \right)
    + \frac{\partial}{\partial y} \left( gE_y(\vec{r},\omega) \right)
    +\frac{\partial}{\partial s} E_s(\vec{r},\omega)
    \right]
    = \frac{\rho(\vec{r},\omega)}{\epsilon_0},
\end{equation}
with $g=1+x/\rho_x(s)$.

Since the source charge and density Eqs.(\ref{eq:Source_charge_density_freq_domain}) and (\ref{eq:Source_current_density_freq_domain}) contain a phase term of $\text{Exp}(i\omega s/v_0)$, it is meaningful to rewrite the fields as
\begin{equation}
    \vec{E}(\vec{r},\omega) =
    \vec{\mathbb{E}}(\vec{r},\omega) e^{\frac{i\omega s}{v_0}},
    \label{eq:Electric_Field_by_amplitude}
\end{equation}
\begin{equation}
    \vec{B}(\vec{r},\omega) =
    \vec{\mathbb{B}}(\vec{r},\omega) e^{\frac{i\omega s}{v_0}}.
    \label{eq:Magnetic_Field_by_amplitude}
\end{equation}
Consequently, the equations for the "amplitude" of fields are
\begin{equation}
    \frac{1}{g} \left[
    \frac{\partial}{\partial y} \left( g\mathbb{E}_s(\vec{r},\omega) \right)
    - \frac{\partial}{\partial s} \mathbb{E}_y(\vec{r},\omega)
    - \frac{i\omega}{v_0} \mathbb{E}_y(\vec{r},\omega)
    \right]
    = i\omega \mathbb{B}_x(\vec{r},\omega),
\end{equation}
\begin{equation}
    \frac{1}{g} \left[
    \frac{\partial}{\partial s} \mathbb{E}_x(\vec{r},\omega)
    + \frac{i\omega}{v_0} \mathbb{E}_x(\vec{r},\omega)
    - \frac{\partial}{\partial x} \left( g\mathbb{E}_s(\vec{r},\omega) \right)
    \right]
    = i\omega \mathbb{B}_y(\vec{r},\omega),
\end{equation}
\begin{equation}
    \frac{\partial}{\partial x} \mathbb{E}_y(\vec{r},\omega)
    - \frac{\partial}{\partial y} \mathbb{E}_x(\vec{r},\omega)
    =i\omega \mathbb{B}_s(\vec{r},\omega),
\end{equation}

\begin{equation}
    \frac{1}{g} \left[
    \frac{\partial}{\partial y} \left( g\mathbb{B}_s(\vec{r},\omega) \right)
    - \frac{\partial}{\partial s} \mathbb{B}_y(\vec{r},\omega)
    - \frac{i\omega}{v_0} \mathbb{B}_y(\vec{r},\omega)
    \right]
    + \frac{i\omega}{c^2} \mathbb{E}_x(\vec{r},\omega)
    = \mu_0 j_{x0}(\vec{r},\omega),
\end{equation}
\begin{equation}
    \frac{1}{g} \left[
    \frac{\partial}{\partial s} \mathbb{B}_x(\vec{r},\omega)
    + \frac{i\omega}{v_0} \mathbb{B}_x(\vec{r},\omega)
    - \frac{\partial}{\partial x} \left( g\mathbb{B}_s(\vec{r},\omega) \right)
    \right]
    + \frac{i\omega}{c^2} \mathbb{E}_y(\vec{r},\omega)
    = \mu_0 j_{y0}(\vec{r},\omega),
\end{equation}
\begin{equation}
    \frac{\partial}{\partial x} \mathbb{B}_y(\vec{r},\omega)
    - \frac{\partial}{\partial y} \mathbb{B}_x(\vec{r},\omega)
    + \frac{i\omega}{c^2} \mathbb{E}_s(\vec{r},\omega)
    =\mu_0 j_{s0}(\vec{r},\omega),
\end{equation}
with $\vec{j}(\vec{r},\omega) = \vec{j}_0(\vec{r},\omega)e^{i\frac{\omega s}{v_0}}$.

\section{Lorentz force}

A test particle lying in the $x$-$z$ plane follows the source particle but with a spatial delay $d=v_0\tau$. 
The position and velocity of the test particle are
\begin{equation}
    \vec{r}_2(s)=x_2\Vec{e}_x + y_2\Vec{e}_y + \Vec{r}_0(s),
\end{equation}
\begin{equation}
    \vec{v}_2 = 
    \frac{d\vec{r}_2(s(t))}{dt}
    =v_0\vec{e}_s,
    \label{eq:Velocity_of_test_particle}
\end{equation}
where $g_2=1+x_2/\rho_x(s_2)$ with $s_2=s-d$. Note that the basis vectors $\vec{e}_x$, $\vec{e}_y$, and $\vec{e}_s$ are those (because they are $s$-dependent in F-S frame) at the position of the test particle $s_2=s-d$.
For simplicity, we assume that both the source and test particles in the beam have no angular divergence, i.e., $x' = y' = 0$.

The Lorentz force law is
\begin{equation}
    \frac{d\vec{p}}{dt} =
    \vec{F}(\vec{r},t) = q \left( \vec{E}(\vec{r},t) + \vec{v} \times \vec{B}(\vec{r},t) \right),
\end{equation}
with $\vec{v} \equiv d\vec{r}/dt$. Since we are switching to the Frenet-Serret coordinate system, the Lorentz force law is modified
\begin{equation}
    \frac{d\vec{p}}{ds} =
    \frac{dt}{ds}
    \frac{d\vec{p}}{dt}
    =
    \frac{dt}{ds} \vec{F}(\vec{r},t(s)).
\end{equation}
Then, we have
\begin{equation}
    \frac{d\vec{p}}{ds} =
    q \left( 
    \frac{dt}{ds} \vec{E}(\vec{r},t(s)) 
    + \frac{d\vec{r}}{ds} \times \vec{B}(\vec{r},t(s)) \right)
    \equiv \vec{\mathcal{F}}/v_0.
\end{equation}

For our case, it is convenient to write it as a function of the coordinates of the source and test particles
\begin{equation}
    \vec{\mathcal{F}}(\vec{R}_2,\vec{R}_1,s) = q_2 v_0 \left( \frac{dt}{ds} \vec{E}(\vec{R}_2,\vec{R}_1,s) + \frac{d\vec{R}_2}{ds} \times \vec{B}(\vec{R}_2,\vec{R}_1,s) \right),
\end{equation}
with $\vec{R}_2=(x_2,y_2,s_2)$, $\vec{R}_1=(x_1,y_1,s_1)$, and
\begin{equation*}
    \frac{d\vec{R}_2}{ds} = \frac{d\vec{R}_2}{dt} \frac{dt}{ds} = g_2 \vec{e}_s.
\end{equation*}

When describing the motion of a particle, $ds/dt$ depends on the local coordinates. When describing the evolution of fields, we can constrain that both $s$ and $t$ are measured by referring to the ideal particle. Consequently, there is $ds/dt=v_0$. Finally, the transverse and longitudinal Lorentz forces are
\begin{equation}
    \mathcal{F}_x (\vec{R}_2,\vec{R}_1,s) =
    q_2 \left( E_x(\vec{R}_2,\vec{R}_1,s) - v_0 g_2 B_y(\vec{R}_2,\vec{R}_1,s)\right),
\end{equation}
\begin{equation}
    \mathcal{F}_y (\vec{R}_2,\vec{R}_1,s) =
    q_2 \left( E_y(\vec{R}_2,\vec{R}_1,s) + v_0 g_2 B_x(\vec{R}_2,\vec{R}_1,s)\right),
\end{equation}
\begin{equation}
    \mathcal{F}_s (\vec{R}_2,\vec{R}_1,s) =
    q_2 g_2 E_s(\vec{R}_2,\vec{R}_1,s).
\end{equation}
with $F_{x,y,s} = \mathcal{F}_{x,y,s}\frac{1}{g_2}$.

\section{Impedance and wake functions}

In principle, solving the Maxwell-Lorentz equations predicts the dynamics of beam and fields simultaneously. But in many cases, a perturbation theory is preferred to simplify the problem. The concepts of impedance and wake functions describe the overall effects of the beam when it traverses an environment. Two approximations (i.e. rigid beam and impulse, referring to~\cite{chao2020lectures}) are necessary to bridge them with the Maxwell-Lorentz equations. The rigid-beam approximation lies in the definition of the source and test charges.

With constraints on the relative positions of the source and test particles, we can obtain an impulse kick by integrating the Lorentz force along the path of the test particle. The constraints between $\vec{R}_2$ and $\vec{R}_1$ are $s_1=v_0t$ and $s_2=v_0t-d$.

The general definitions of wake potentials are as follows.
\begin{equation}
    w_s(\vec{r}_2,\vec{r}_1; d) = 
    -\frac{1}{q_1q_2}
    \int_{-\infty}^{\infty} ds 
    \vec{e}_s \cdot
    \left. \vec{\mathcal{F}}(\vec{R}_2,\vec{R}_1,s) \right | _{s_2=s_1-d},
    \label{eq:Wake_function_lontigudinal}
\end{equation}
\begin{equation}
    w_x(\vec{r}_2,\vec{r}_1; d) = 
    \frac{1}{q_1q_2}
    \int_{-\infty}^{\infty} ds
    \vec{e}_x \cdot
    \left. \vec{\mathcal{F}}(\vec{R}_2,\vec{R}_1,s) \right | _{s_2=s_1-d},
    \label{eq:Wake_function_horizontal}
\end{equation}
\begin{equation}
    w_y(\vec{r}_2,\vec{r}_1; d) = 
    \frac{1}{q_1q_2}
    \int_{-\infty}^{\infty} ds
    \vec{e}_y \cdot
    \left. \vec{\mathcal{F}}(\vec{R}_2,\vec{R}_1,t) \right | _{s_2=s_1-d},
    \label{eq:Wake_function_vertical}
\end{equation}
with $(\vec{e}_x,\vec{e}_y,\vec{e}_s)$ forming the right-handed orthonormal basis. The minus sign in $w_s$ is a choice of convention. The F-S frame is one popular choice in accelerator physics. The variables $\vec{r}_2=(x_2,y_2)$ and $\vec{r}_1=(x_1,y_1)$ represent the transverse coordinates of the test and source particles, respectively. Using fields, the explicit formulations of wake functions are
\begin{equation}
    w_s(\vec{r}_2,\vec{r}_1; d) = 
    -\frac{1}{q_1}
    \int_{-\infty}^{\infty} ds g_2 
    \left. E_s(\vec{R}_2,\vec{R}_1,s) \right | _{s_2=s_1-d},
    \label{eq:Wake_function_lontigudinal_simplified}
\end{equation}
\begin{equation}
    \begin{split}
        w_x(\vec{r}_2,\vec{r}_1; d) 
        & = 
        \frac{1}{q_1q_2}
        \int_{-\infty}^{\infty} ds
        \left. \mathcal{F}_x(\vec{R}_2,\vec{R}_1,s) \right | _{s_2=s_1-d} \\
        & = \frac{1}{q_1}
        \int_{-\infty}^{\infty} ds g_2
        \left. \left(E_x(\vec{R}_2,\vec{R}_1,s) - v_0B_y(\vec{R}_2,\vec{R}_1,s)\right) \right | _{s_2=s_1-d}
        ,
    \end{split}
    \label{eq:Wake_function_horizontal_FS}
\end{equation}
\begin{equation}
    \begin{split}
        w_y(\vec{r}_2,\vec{r}_1; d) 
        & = 
        \frac{1}{q_1q_2}
        \int_{-\infty}^{\infty} ds
        \left. \mathcal{F}_y(\vec{R}_2,\vec{R}_1,s) \right | _{s_2=s_1-d} \\
        & = \frac{1}{q_1}
        \int_{-\infty}^{\infty} ds g_2
        \left. \left(E_y(\vec{R}_2,\vec{R}_1,s) + v_0B_x(\vec{R}_2,\vec{R}_1,s)\right) \right | _{s_2=s_1-d}
        ,
    \end{split}
    \label{eq:Wake_function_vertical_FS}
\end{equation}

The constraints between $\vec{R}_2$ and $\vec{R}_1$ further simplify the forms of wake potentials and eventually introduce the impedance, which is the Fourier transform of wake potentials. Utilizing the "amplitude" defined in Eqs.(\ref{eq:Electric_Field_by_amplitude}) and (\ref{eq:Magnetic_Field_by_amplitude}), the wake potentials can be formulated as
\begin{equation}
    w_s(\vec{r}_2,\vec{r}_1; d) = 
    -\frac{1}{2\pi q_1}
    \int_{-\infty}^{\infty} ds
    \int_{-\infty}^{\infty} d\omega g_2
    \left. \mathbb{E}_s(\vec{R}_2,\vec{r}_1,\omega)
    e^{\frac{i\omega s_2}{v_0}-i\omega t}
    \right | _{\substack{s_1=v_0t, \\ s_2=v_0t-d}},
    \label{eq:Wake_function_lontigudinal_simplified}
\end{equation}
\begin{equation}
    \begin{split}
        w_x(\vec{r}_2,\vec{r}_1; d) 
        = & \frac{1}{2\pi q_1}
        \int_{-\infty}^{\infty} ds
        \int_{-\infty}^{\infty} d\omega \\
        & \times g_2 \left. \left(\mathbb{E}_x(\vec{R}_2,\vec{r}_1,\omega) - v_0\mathbb{B}_y(\vec{R}_2,\vec{r}_1,\omega)\right) 
        e^{\frac{i\omega s_2}{v_0}-i\omega t}
        \right | _{\substack{s_1=v_0t, \\ s_2=v_0t-d}}
        ,
    \end{split}
    \label{eq:Wake_function_horizontal_FS_simplified}
\end{equation}
\begin{equation}
    \begin{split}
        w_y(\vec{r}_2,\vec{r}_1; d) 
        = & \frac{1}{2\pi q_1}
        \int_{-\infty}^{\infty} ds
        \int_{-\infty}^{\infty} d\omega \\
        & \times g_2 \left. \left(\mathbb{E}_y(\vec{R}_2,\vec{r}_1,\omega) + v_0\mathbb{B}_x(\vec{R}_2,\vec{r}_1,\omega)\right) 
        e^{\frac{i\omega s_2}{v_0}-i\omega t}
        \right | _{\substack{s_1=v_0t, \\ s_2=v_0t-d}}
        .
    \end{split}
    \label{eq:Wake_function_vertical_FS_simplified}
\end{equation}
Note that the longitudinal coordinate $s_1$ of the source charge is absorbed thanks to the $\delta$-function distribution $\delta(s_1-v_0t)$ of charge density. Finally, we find that the wake potentials in their general forms are functions of the transverse coordinates ($x_1,y_1,x_2,y_2$) of the source and test particles, and the distance between them $d$.

Equations (\ref{eq:Wake_function_lontigudinal_simplified}-\ref{eq:Wake_function_vertical_FS_simplified}) are further simplified as
\begin{equation}
    w_s(\vec{r}_2,\vec{r}_1; d) = 
    -\frac{1}{2\pi q_1}
    \int_{-\infty}^{\infty} ds
    \int_{-\infty}^{\infty} d\omega g_2
    \mathbb{E}_s(\vec{R}_2,\vec{r}_1,\omega)
    e^{-\frac{i\omega d}{v_0}}
    ,
    \label{eq:Wake_function_lontigudinal_simplified_2}
\end{equation}
\begin{equation}
    \begin{split}
        w_x(\vec{r}_2,\vec{r}_1; d) 
        = & \frac{1}{2\pi q_1}
        \int_{-\infty}^{\infty} ds
        \int_{-\infty}^{\infty} d\omega \\
        & \times g_2 \left(\mathbb{E}_x(\vec{R}_2,\vec{r}_1,\omega) - v_0\mathbb{B}_y(\vec{R}_2,\vec{r}_1,\omega)\right) 
        e^{-\frac{i\omega d}{v_0}}
        ,
    \end{split}
    \label{eq:Wake_function_horizontal_FS_simplified_2}
\end{equation}
\begin{equation}
    \begin{split}
        w_y(\vec{r}_2,\vec{r}_1; d) 
        = & \frac{1}{2\pi q_1}
        \int_{-\infty}^{\infty} ds
        \int_{-\infty}^{\infty} d\omega \\
        & \times g_2 \left(\mathbb{E}_y(\vec{R}_2,\vec{r}_1,\omega) + v_0\mathbb{B}_x(\vec{R}_2,\vec{r}_1,\omega)\right) 
        e^{-\frac{i\omega d}{v_0}}
        .
    \end{split}
    \label{eq:Wake_function_vertical_FS_simplified_2}
\end{equation}
Now, the wake functions are recognized as Fourier transforms of impedance:
\begin{equation}
    w_s(\vec{r}_2,\vec{r}_1; d) =
    \frac{1}{2\pi}
    \int_{-\infty}^{\infty} d\omega
    Z_s(\vec{r}_2,\vec{r}_1; \omega)
    e^{-\frac{i\omega d}{v_0}}
    ,
    \label{eq:Wake_impedance_pair_longitudinal}
\end{equation}
\begin{equation}
    w_x(\vec{r}_2,\vec{r}_1; d) =
    \frac{1}{2\pi\kappa}
    \int_{-\infty}^{\infty} d\omega
    Z_x(\vec{r}_2,\vec{r}_1; \omega)
    e^{-\frac{i\omega d}{v_0}}
    ,
    \label{eq:Wake_impedance_pair_horizontal}
\end{equation}
\begin{equation}
    w_y(\vec{r}_2,\vec{r}_1; d) =
    \frac{1}{2\pi\kappa}
    \int_{-\infty}^{\infty} d\omega
    Z_y(\vec{r}_2,\vec{r}_1; \omega)
    e^{-\frac{i\omega d}{v_0}}
    ,
    \label{eq:Wake_impedance_pair_vertial}
\end{equation}
with $\kappa = i/(v_0/c)$. The parameter $\kappa$ is a conventional choice. Mathematically, it is natural to choose $\kappa=1$. Then, the impedance and wake functions in three directions are simply Fourier pairs. Reversely, the impedances are expressed in terms of wake potentials as
\begin{equation}
    Z_s(\vec{r}_2,\vec{r}_1; \omega) =
    \int_{-\infty}^{\infty} d\tau
    w_s(\vec{r}_2,\vec{r}_1; \tau)
    e^{i\omega \tau}
    ,
    \label{eq:Wake_impedance_pair_longitudinal_reverse}
\end{equation}
\begin{equation}
    Z_x(\vec{r}_2,\vec{r}_1; \omega) =
    \kappa \int_{-\infty}^{\infty} d\tau
    w_x(\vec{r}_2,\vec{r}_1; \tau)
    e^{i\omega \tau}
    ,
    \label{eq:Wake_impedance_pair_horizontal_reverse}
\end{equation}
\begin{equation}
    Z_y(\vec{r}_2,\vec{r}_1; \omega) =
    \kappa \int_{-\infty}^{\infty} d\tau
    w_y(\vec{r}_2,\vec{r}_1; \tau)
    e^{i\omega \tau}
    ,
    \label{eq:Wake_impedance_pair_horizontal_reverse}
\end{equation}
with $\tau\equiv d/v_0$.

Comparing Eqs. (\ref{eq:Wake_impedance_pair_longitudinal}-\ref{eq:Wake_impedance_pair_vertial}) with Eqs. (\ref{eq:Wake_function_lontigudinal_simplified_2}-\ref{eq:Wake_function_vertical_FS_simplified_2}), we can obtain the impedances in terms of "amplitude" fields:
\begin{equation}
    Z_s(\vec{r}_2,\vec{r}_1; \omega)
    = -\frac{1}{q_1}
    \int_{-\infty}^{\infty} ds g_2
    \mathbb{E}_s(\vec{R}_2,\vec{r}_1,\omega),
    \label{eq:Impedance_by_fields_longitudinal}
\end{equation}
\begin{equation}
    Z_x(\vec{r}_2,\vec{r}_1; \omega)
    = \frac{\kappa}{q_1}
    \int_{-\infty}^{\infty} ds g_2
    \left(\mathbb{E}_x(\vec{R}_2,\vec{r}_1,\omega) - v_0\mathbb{B}_y(\vec{R}_2,\vec{r}_1,\omega)\right),
    \label{eq:Impedance_by_fields_horizontal}
\end{equation}
\begin{equation}
    Z_y(\vec{r}_2,\vec{r}_1; \omega)
    = \frac{\kappa}{q_1}
    \int_{-\infty}^{\infty} ds g_2
    \left(\mathbb{E}_y(\vec{R}_2,\vec{r}_1,\omega) + v_0\mathbb{B}_x(\vec{R}_2,\vec{r}_1,\omega)\right).
    \label{eq:Impedance_by_fields_vertical}
\end{equation}

\section{Discussion}

The transverse and longitudinal impedances (and their corresponding wake potentials) are correlated. This is the nature of the Lorentz force. Formulating the correlations are not trivial in the F-S system. An attempt was presented in~\cite{zhou2023generalized}.


\bibliography{impedance}

\end{document}